\documentstyle[preprint,eqsecnum,aps]{revtex}

\begin{document}
\draft
\preprint{SNUTP-94-10}

\title{Abelian Chern-Simons field theory \\
and anyon equation on a cylinder}

\author { Kyung-Hyun Cho and Chaiho Rim\\
Department of Physics\\
Chonbuk National University\\
Chonju, 560-756,Korea}

\date {January 26, 1994}
\maketitle

\begin{abstract}

We present the anyon equation on a cylinder
and in an infinite potential wall
from the abelian Chern-Simons
theory  coupled to non-relativistic
matter field by obtaining the effective hamiltonian
through the canonical transformation
method used for the theory
on a plane and on a torus.
We also give the periodic property of the theory
on the cylinder.

\end{abstract}


\narrowtext
\pagebreak

\section{Introduction}

It is well known
that the Chern-Simons field theory \cite{deser,hagen} provides
an effective description of the statistical transmutation
\onlinecite{goldhaber,semenoff,matsuyama,swanson-cho}
whose effect appears
as the Aharonov-Bohm potential in the many-body equation
and give rise to the anyonic property
\cite{leinaas}.
For the non-relativistic case, this is explicitly demonstrated
by obtaining the Schr\"odinger equation
on an infinite plane \cite{jackiw,kimlee}
and on compact spaces
\onlinecite{randjbar,hoso,rim-torus}
from the
Chern-Simons gauge theory coupled to non-relativistic
matter-field.

The derivation of the
anyon equation on the compact space from the
Chern-Simons field theory is not straight-forward
due to the zero-modes and the boundary condition
of the wavefunction is not well-understood.
We have re-analyzed
\cite{rim-torus}
the Chern-Simons theory
on a torus extending the canonical
transformation method used in the plane \cite{swanson-cho}.
The canonical transformation decouples the gauge fields
from the matter sector and
leaves the effective hamiltonian in terms of
matter fields only. This hamiltonian system
is equivalent
to the one where Gauss constraint is
solved. In addition, the system has the manifest
translation invariance along the
non-contractible loops.
 From this we could
obtain the many-body Schr\"odinger equation
with simple periodic (not quasi-periodic)
boundary condition.
It does not exclude the possibility
of the quasi-periodic boundary condition.
Instead, depending on the definition of the wavefunction,
one can have the quasi-periodic condition
due to the vacuum degeneracy
of the gauge sector \cite{polychro} as presented previously
\onlinecite{randjbar}.

The canonical transformation avoids the
ambiguity appearing  in solving the Gauss constraint classically
and quantizing the zero-modes and the matter fields afterwards.
It is worthwhile to apply this method to other spaces
with different topology
and obtain its effective theory and many-body quantum mechanics.
In this paper, we will consider
the Chern-Simons theory
on a cylinder and inside an infinite potential wall.
The cylinder is compact in the one direction
and infinite in the other. Therefore,
it will share some of  the properties on torus
and  on plane.
In addition, it will be interesting
how the results change if one introduces
the infinite potential wall.

This paper is organized as follows.
In section II, we  consider
the quantization of abelian pure Chern-Simons gauge field
on a strip (with the cylindrical boundary condition)
and obtain an effective hamiltonian
of matter field in the
fundamental domain of the cylinder
by defining new fields through a canonical
transformation.
In section III,
we construct the many-body
Schr\"odinger equation on the strip.
The equation possesses
the Aharonov-Bohm potential
which is adapted to the cylindrical boundary condition
and is responsible for the
statistical  transmutation.
In section IV, we consider
the periodic property of the fields and
the boundary condition of the wavefunction.
In section V,
we comment on the theory
inside a (rectangular) hard wall.
Section V is the summary of the result and discussion.

\section{Hamiltonian  on a Strip}

Let us consider the system in the fundamental
domain of the cylinder,
$D_{0}$ ($0 \le x^1 < L_1$
and $-\infty  < x^2 < \infty $).
The lagrangian density of abelian Chern-Simons
gauge theory coupled to non-relativistic matter field
is given as
\begin{equation}
{\cal L}= {\mu \over 2}
\epsilon^{\mu\nu\rho} a_\mu \partial_\nu a_\rho
+ \psi^+ i\hbar D_0 \psi
- {\hbar^2 \over 2m} |\vec D \psi|^2\,,
\label{lagcyl}
\end{equation}
where
$i\hbar D_\mu \psi
=  (i \hbar \partial_\mu
- {e\over c } a_\mu ) \psi
$.
We will choose the matter field
$ \psi $ as a fermion-field for definiteness.
The analysis goes same with the bosonic case.

The gauge field does not propagate
and the hamiltonian system becomes a constraint one.
The phase space variables of the gauge fields
are  $a_1$ and $\pi_1$,
where
\begin{equation}
\pi_1 = {\partial {\cal L} \over \partial \dot a_1}
= {\mu \over c} a_2\,,
\end{equation}
and the gauge-field itself constitutes the phase space.
The hamiltonian density of the system is written as
\begin{equation}
{\cal H}=
a_0 (\mu b + {e\over c}J_0)
+ {1\over 2m} |i\hbar D_i \psi|^2\,,
\label{hamil}
\end{equation}
where $J_0 = \psi^+ \psi$ and
$b = -(\partial_1a_2 -\partial_2a_1)$.
The operators satisfy the equal-time
(anti-) commutation  relations;
$$ \{\psi(x), \psi^+(y)\} =  \delta (\vec x - \vec y) $$
$$ [J_0(x), \psi^+(y)] = \psi^+(x) \delta (\vec x - \vec y) $$
\begin{equation}
[a_1(x), a_2(y)] = i {\hbar \over \mu}
\delta(\vec x - \vec y)\,.
\label{commfield}
\end{equation}
Here $x$
denotes the three vector $(ct, \vec x )$ and
$\vec x =(x^1, x^2) = (-x_1, -x_2)$.

$a_0(x)$ is treated as a Lagrange
multiplier and  commutes with $a_i$'s.
The constraint is given as the Gauss law,
$\Gamma \cong 0$ where
\begin{equation}
\Gamma \equiv   b + {e \over \mu c} J_0
\end{equation}
is the gauge generator.
A physical state should be annihilated by $\Gamma$,
$\Gamma |{\rm phys}> = 0$.

On the strip  with (period) length $L_1$
(the fundamental domain of the cylinder),
the gauge operators are decomposed as
\begin{equation}
a_i(x) =
{1 \over \sqrt {2\pi L_1 }}
\sum_{n_1 }\int d p_2\,
{1 \over { p^0}}
[ ( {p_i \over \mu} g(\vec p)
+ i{\epsilon_{ij} p_j \over 2  } h (\vec p))e^{-ip\cdot x}
+ c.c]_{p^0 = |\vec p|}\,,
\label{aicyl}
\end{equation}
where $n_1 = { L_1 p_1 \over 2 \pi}$ is an integer.
$g$ and $h$ satisfy
the commutation relation,
$$ [g(\vec p), h^+(\vec q)]
= \hbar \delta_{p_1, q_1} \delta(p_2 -q_2)
$$
\begin{equation}
[g(\vec p), h(\vec q)]
=[g(\vec p), g^+(\vec q)]
= [h(\vec  p ), h^+(\vec q)]
=0\,.
\label{ghcomm}
\end{equation}
$a_0(x)$ commutes with other fields and
is expressed  as
\begin{equation}
a_0(x)
= \lambda  + {1 \over \sqrt {2\pi L_1}}
\sum_{n_1 } \int d p_2 \,
[ {g(\vec p)\over \mu}  e^{-ip\cdot x}
+ c.c]_{p^0 = |\vec p|}\,,
\label{a0cyl}
\end{equation}
where $\lambda$ is a constant (time-independent) Lagrange multiplier.
The statistical magnetic field $b$ is given as
\begin{equation}
b(x) = {1 \over \sqrt {2\pi L_1}}
\sum_{n_1 } \int  d p_2\,
[{p^0 \over 2  } h(\vec p)  e^{-ip\cdot x}
+ c.c]_{p^0 = |\vec p|}\,.
\label{bcyl}
\end{equation}

The above mode expansions are constructed
such that $[ a_0 , a_i]=0$, and
the Lorentz gauge fixing condition
is satisfied  identically,
$\partial_\mu a^\mu = 0$.
There is still left a residual
gauge  degree of freedom with the
Lorentz gauge fixing condition maintained:
$a_\mu \rightarrow a_\mu + \partial_\mu  \Lambda$
where $\partial^2 \Lambda=0$.
This merely redefines $g(\vec p)$ in
Eqs.~ (\ref{aicyl}, \ref{a0cyl}) satisfying the same commutation
relations in Eq. (\ref{ghcomm}).
In our analysis,
we do not need the specific form of $g(\vec p)$ and therefore,
we can proceed in the residual gauge independent manner.

The mode decomposition of the gauge field $a_i(x)$ in
Eq. (\ref{aicyl}) contains the zero-modes ($n_1=0$),
which give rise to the mean field contribution
to the $b$-field in Eq. (\ref{bcyl}).
This is in contrast with the case on a torus, where
we have to introduce the mean field mode in the canonical
transformation formalism.
On the other hand, the natural occurence of the
mean-field contribution is similar to
the case on the infinite plane.
This feature leads us to follow the canonical formalism
used in the infinite plane case.
We will consider two procedures adopted for the infinite
plane and on the torus sepreately and compare the
distinct features.

Let us try first
the canonical transformation
following the method \cite{swanson-cho} used
in the field theory on a plane
to get the effective hamiltonian
and decouple the gauge field from the physical Hilbert space.
The canonical transformation operator is given as
\begin{equation}
V_1(t)= \exp i
{e\over \hbar c}
\int\! \int d\vec x d \vec y
J_0^{(1)}(\vec x,t)
G_{\rm cyl}(\vec x, \vec y) \partial_k a_k^{(1)}(\vec y, t)
\,.
\label{V_1}
\end{equation}
$G_{\rm cyl}$ is the periodic (single-valued on the cylinder)
Green's function   satisfying
\begin{equation}
\nabla^2 G_{\rm cyl}(\vec x, \vec y)
=\delta( \vec x -\vec y )
\nonumber\\
\end{equation}
\begin{equation}
G_{\rm cyl}(\vec x, \vec y)
= G_{\rm cyl}(\vec x, \vec y + L_1 \hat e_1)\,; \quad
G_{\rm cyl}(\vec x, \vec y) = G_{\rm cyl} (\vec y, \vec x )
\end{equation}
whose explicit form is given as
\begin{equation}
G_{\rm cyl}(\vec x, \vec y)
= {1\over 4\pi} \ln
\left| \sin \pi z \right|^2
\label{Gcyl}
\end{equation}
where
$z= [(x^1-y^1) + i (x^2 - y^2)]/L_1$.
We regularize the Green's function as
\begin{equation}
\epsilon_{ij} \partial_{x^j} G_{\rm cyl}(\vec x, \vec y)
\left.\right|_{\vec x = \vec y}  =0\,.
\label{greenreg}
\end{equation}

The original fields are given  in terms  of the new fields as
\begin{eqnarray}
\psi(x) & =& V_1(t) \psi^{(1)}(x) V_1^+(t)
\nonumber\\
& =& \psi^{(1)}(x)
\exp -{ i e\over \hbar c}
\int \! d \vec y
G_{\rm cyl} (\vec x, \vec y)  \partial_{y^i} a_i^{(1)} (y)\,,
\nonumber\\
a_i(x) &=&
V_1(t) a_i^{(1)}(x) V_1^+(t)
\nonumber\\
&=& a_i^{(1)}(x) -
{e \over \mu c}
\epsilon_{ij} \partial_{x^j}
\int \! d\vec y {G}_{\rm cyl}(\vec x, \vec y) J_0^{(1)} (y)\,,
\nonumber\\
\label{field1def}
\end{eqnarray}
where $J_0(x)= J_0^{(1)}(x)$.
The new fields satisfy the same form
of the commutation relations
given in Eq. (\ref{commfield}).

Expressing the  hamiltonian density ${\cal H}$
in terms of  the new fields, we have
\begin{equation}
{\cal H}= \mu a_0 b^{(1)}
+ {1 \over 2m}  \left|
(i \hbar \partial_i - {e\over c} a_i^{({\rm eff})} )
\psi^{(1)} \right|^2\,.
\label{hamil1}
\end{equation}
Implicit ordering of $\psi^{(1)+}$ to the left of $a_i^{({\rm eff})}$
is assumed.
Here $b^{(1)} = - (\partial_1 a_2^{(1)} - \partial_2 a_1^{(1)}) $
is the gauge generator
since $ b^{(1)} = b + {e \over \mu c} J_0 = \Gamma$
from Eq. (\ref{field1def}).
$a_i^{({\rm eff})}$ is an effective gauge-like term,
\begin{equation}
a_i^{({\rm eff})}(x)
= \epsilon_{ij} \partial_{x^j}
\int \! d \vec y {G}_{\rm cyl}(\vec x, \vec y)
(b^{(1)}(y) -{e \over \mu c} J_0(y))
\,. \label{aeff}
\end{equation}
This is obtained by noting that (see Eq. (\ref{aicyl}))
\begin{equation}
a_i^{(1)}(x)
= \epsilon_{ij} \partial_{x^j}
\int \! d \vec y {G}_{\rm cyl}(\vec x, \vec y) b^{(1)}(x)
+ \partial_{x^i} \int \! d \vec y {G}_{\rm cyl}(\vec x, \vec y)
\partial_{y^j} a_j^{(1)} (y) \,.
\label{a1i}
\end{equation}
We emphasize that the readers should not confuse the gauge field
$ a_i^{(1)}(x)$ with $a_i^{({\rm eff})}(x)$.
$ a_i^{(1)}(x)$'s are satisfying the same
commutation  relation as that given in Eq. (\ref{commfield}).
However, the commutation
relation between  $a_i^{({\rm eff})}(x)$'s is given as
\begin{equation}
[a_1^{({\rm eff})}(x), a_2^{({\rm eff})}(y)]
= 0\,.
\end{equation}

The gauge transformation operator of
this system is given as
\begin{equation}
B \equiv i {\mu \over \hbar}
\int \! d \vec x \, b^{(1)}(x) \Lambda(x)\,,
\end{equation}
where $\Lambda(x)$ is an arbitrary real periodic function.
$a_i^{(1)}(x)$, $a_i(x)$ and
$\psi(x)$ are  gauge-dependent since
$$ [B, a_i^{(1)}(x)] = \partial_i \Lambda(x)\,,
\quad
[B, a_i(x)] = \partial_i \Lambda(x)\,,
$$
\begin{equation}
[B, \psi(x)] = {-ie\over \hbar c} \Lambda (x) \psi(x)\,,
\end{equation}
whereas $\psi^{(1)}(x)$ is gauge invariant
\begin{equation}
[B, \psi^{(1)}(x)] =0\,.
\end{equation}
Since the gauge generator $b^{(1)}(x)$ commutes
with any other operators in  ${\cal H}$
and since we consider the gauge
invariant  operators and physical states only,
we may drop  $b^{(1)}(x)$ completely in $\cal H$.
Therefore, the gauge fields  disappear
from the hamiltonian density,
\begin{equation}
{\cal H}(x)
=  {1 \over 2m} \left| (i \hbar \partial_i -
{e\over c} A_i^{[1]}(x) )\psi^{(1)}(x) \right|^2\,,
\label{hamilN}
\end{equation}
where
\begin{equation}
A_i^{[1]}(x)= -{e \over \mu c}
\epsilon_{ij} \partial_{x^j}
\int \! d \vec y {G}_{\rm cyl}(\vec x, \vec y) J_0(y)\,.
\label{A1i}
\end{equation}
The effective hamiltonian density
is the desired form, which amounts to solving the Gauss
constraint
and the newly  defined fields enable us to construct  the
physical state by simply applying operators consisting of
$\psi^{(1)+}$
on the vacuum $|0>$
which satisfies  $b^{(1)}|0>=J_0|0>=0$.

The canonical transformation is successful
in obtaining the effective interaction.
One may wonder if there is any subtleties observed
in the canonical formlism
on a compact space such as on a torus
since a cylinder has  a finite dimension in one direction
and an infinite on the other.
We note that one of the subtleties arises on the
boundary condition of the Green's function on the compact space.
To see this more explicitly, we may re-investigate the
new gauge field $a_i^{(1)}(x)$ given in Eq. (\ref{field1def}).
To obtain this, we discarded the boundary term
considering that it vanishes. It really vanishes
at the edge of the strip since the Green's function
is given in Eq. (\ref{Gcyl}). On the other hand,
at infinity of the strip $G_{\rm cyl}$
increases linearly with the coordinate, but one still regards the
contributions at $\pm \infty$ cancel each other.

One may avoid the linear growth of the Green's function
by removing the asymptotically increasing term and define
another periodic Green's function
\begin{equation}
G_p (\vec x, \vec y)
= {1\over 4\pi} \ln
\left| \sin \pi z \right|^2
- {1 \over 2}|x_2 - y_2|\,,
\label{G_p}
\end{equation}
satisfying
\begin{equation}
\nabla^2 G_p (\vec x, \vec y)
= (\delta(x^1-y^1) - {1\over L_1})\delta(x^2-y^2).
\nonumber\\
\end{equation}
We can consider another canonical operator $V_2(t)$
in analogy as in the torus \cite{rim-torus}
using the Green's function $G_p(\vec x, \vec y)$,
\begin{equation}
V_2(t)= \exp i
{e\over \hbar c}
\int\! \int d\vec x d \vec y
J_0^{(2)}(\vec x,t)
\left[ g_2(\vec x, \vec  y)
{\theta_2^{(2)}(\vec y, t)}
+ G_p(\vec x, \vec y)
\partial_k a_{k}^{(2)}(\vec y, t) \right] \,.
\label{V_2}
\end{equation}
$\theta_i$ is the analogue of the zero-mode on a torus,
\begin{eqnarray}
\theta_1(x) &=& {1 \over 2 \sqrt{ 2 \pi L_1}}
\int d p_2 [ i h(\vec p ) e^{- i (p \cdot x )}
+ c.c]|_{p_0 = |p_2|, p_1 = 0}
\nonumber\\
\theta_2(x) &= &{1 \over \mu \sqrt{ 2 \pi L_1}}
\int d p_2 [g(\vec p ) e^{- i ( p \cdot x )}
+ c.c]|_{p_0 = |p_2|, p_1 = 0}
\label{thetamode}
\end{eqnarray}
which satisfy the equal-time commutation relation
\begin{equation}
[\theta_1(x), \theta_2(y)] = {i\hbar \over \mu L_1} \delta(x_2 - y_2)\,.
\end{equation}
$g_2(\vec x, \vec y)$ is a function satisfying
$\partial_2 g_2 (\vec x, \vec y) = \delta (x_2 - y_2)$,
whose explicit form we may choose as
\begin{equation}
g_2(\vec x, \vec y) = {1\over 2} \epsilon(x^2 - y^2)\,.
\end{equation}
The canonical operator $V_2(t)$ reduces to $V_1(t)$ if we neglect
the boundary term: $V_2(t)$ is an improved operator.

The relation of the new fields with the original ones is given as
\begin{eqnarray}
\psi(x) & =& V_2(t) \psi^{(2)}(x) V_2^+(t)
\nonumber\\
& =& \psi^{(2)}(x)
\exp -{ i e\over \hbar c}
\int \! d \vec y
\{ g_2(\vec x, \vec y) {\theta_2(y) }
+ G_p(\vec x, \vec y)  \partial_{y^i} a_i^{(2)} (y) \}\,,
\nonumber\\
a_i(x) &=&
V(t) a_i^{(2)}(x) V^+(t)
\nonumber\\
&=& a_i^{(2)}(x) -
{e \over \mu c}
\epsilon_{ij} \partial_{x^j}
\int \! d\vec y {G}_{\rm cyl}(\vec x, \vec y) J_0^{(2)} (y)\,,
\nonumber\\
\label{field2def}
\end{eqnarray}
where $J_0(x)= J_0^{(2)}(x)$ and
$b^{(2)} = -(\partial_1 a_2^{(2)} - \partial_2 a_1^{(2)})$.
Expressing the  hamiltonian density ${\cal H}$
in terms of  the new fields, we have
\begin{equation}
{\cal H}= \mu a_0 b^{(2)}
+ {1 \over 2m}  \left|
(i \hbar \partial_i - {e\over c} a_i^{({\rm eff})} )
\psi^{(2)} \right|^2\,.
\label{hamil2}
\end{equation}
Implicit ordering of $\psi^{(2)+}$ to the left of $a_i^{({\rm eff})}$
is assumed.
Here $b^{(2)}=\Gamma =b^{(1)}$
is the gauge generator
since $ b^{(2)} = b + {e \over \mu c} J_0 $ from Eq. (\ref{field2def}).
$a_i^{({\rm eff})}$ is the same effective gauge-like term
given in Eq. (\ref{aeff}).
The effective hamiltonian does reduce  to the same form
as obtained from the canonical transformation using $V_1$.
Restricting the system to the physical Hilbert space
we decouple the gauge field
from the system and regain the hamiltonian given
in Eq. (\ref{hamilN}).


\section{Many-body quantum mechanics}

The many-body quantum mechanics can be obtained from the
field theory using the Heisenberg equation of motion.
Let us define the
$N$-particle wavefunction as
\begin{equation}
\Phi (1, \cdots, N)
\equiv < 0|\psi^{(2)} (x^{(1)}) \cdots \psi^{(2)} (x^{(N)}) |N > \,.
\label{Nwave}
\end{equation}
We assume that all the coordinates
of the particles lie in a fundamental domain.
$|0 >$ is the vacuum which satisfies $J_0|0 > = 0$
and $|N >$ is the $N$-body
Heisenberg state vector.
The wavefunction is gauge invariant
since
$\psi^{(2)}(x)$ is gauge invariant.
Then the
Schr\"odinger equation is given as
\begin{equation}
i\hbar {\partial\Phi \over \partial t} (1, \cdots, N)
= \sum_{p=1}^N < 0 | \psi^{(2)} (x^{(1)})
\cdots i\hbar {\partial\psi^{(2)} (x^{(p)})  \over \partial
t} \cdots \psi^{(2)} (x^{(N)}) |N > \,.
\end{equation}
The time evolution of the matter-field  operator
is given as the Heisenberg equation  of motion,
$$
i\hbar {\partial\psi^{(2)} (x)
\over \partial t} = [\psi^{(2)} (x), H]$$
$$
= {1 \over 2m} \left[ \right.
(i\hbar\partial_{x^i} - {e \over e} A_i^{[1]} (x))^2 +
({e \over c})^2 \int d\vec y \psi^{(2)+} (y)
K_i(y,x) K_i(y,x)\psi^{(2)} (y) $$
$$ - ({e \over c}) \int d\vec y \psi^{(2)+} (y) (i\hbar
\stackrel{\leftarrow}\partial_{y^i} -
({e \over c}) A_i^{[1]}  (y)) K_i(y,x)
\psi^{(2)} (y) $$
\begin{equation}
- ({e \over c}) \int d\vec y \psi^{(2)+} (y))K_i(y,x)
 (i\hbar \vec \partial_{y^i} -
({e \over c}) A_i^{[1]}  (y)) \psi^{(2)} (y)
\left. \right] \psi^{(2)} (x)\,,
\label{NSch}
\end{equation}
where
\begin{equation}
K_i(x,y) = - {e \over \mu c} \epsilon_{ij} \partial_{x^j}
G_{\rm cyl}(\vec x, \vec y)\,.
\label{Kgreencyl}
\end{equation}
$G_{\rm cyl}(\vec x, \vec y)$ is
given in Eq.~(\ref{Gcyl}).
We put $\psi^{(2)+}$ and $A_i^{[1]}$
to the far left in each term such that the
operators vanish when they act on
the vacuum $<0 |$ and used the identity
$$
[\psi^{(2)} (x), A_i^{[1]} (y)]  = K_i (y,x) \psi^{(2)} (x)\,.
$$

The  one-particle wavefunction satisfies the Schr\"odinger equation
\begin{equation}
i\hbar {\partial\Phi \over \partial t} (x)
= < 0 | i\hbar {\partial\psi^{(2)}
(x) \over \partial t} |1 >
= - {\hbar^2  \over 2m} \nabla ^2 \, \Phi(x)\,.
\end{equation}
The equation does reduce to the free one.
For the two-body case, we have
\begin{eqnarray}
i\hbar {\partial\Phi \over \partial t} (1,2)
&=& < 0 | \psi^{(2)} (x^{(1)})
i\hbar {\partial\psi^{(2)} (x^{(2)}) \over \partial t} +
i\hbar {\partial\psi^{(2)} (x^{(1)})
\over \partial t} \psi^{(2)} (x^{(2)}) |2 >
\nonumber\\
&=& {1 \over 2m} \{(i\hbar \partial_i^{(1)}
- {e  \over c} {\cal A}_i (1,2))^2 +
(i\hbar \partial_i^{(2)} - {e  \over c}
{\cal A}_i (2,1))^2 \} \Phi(1,2)\,,
\end{eqnarray}
where
\begin{equation}
{\cal A}_i (p,r) =  - {e \over \mu c}
\epsilon_{ik} \partial_k^{(p)} G_{\rm cyl}(p,r)\,,
\end{equation}
is the Aharonov-Bohm potential.
In general for the $N$-particle case, we have
\begin{equation}
i\hbar {\partial\Phi \over \partial t} (1, \cdots, N)
= {1 \over 2m} \sum^N_{p=1}
\left\{
i\hbar \partial_i^{(p)} - {e  \over c}
\sum^N_{r=1(\ne p)} {\cal A}_i (p,r) \right\}^2
\Phi(1, \cdots, N)\,.
\label{NSchp}
\end{equation}

The Aharonov-Bohm gauge potential can be
transformed away through the singular gauge transformation
as in the infinite plane case.
Explicitly,  the gauge potential
becomes
$$ {\cal A}_i (p,r)
= - \partial_i^{(p)}
\left[ {ie    \over 4\pi \mu c}
\ln {\sin(\pi z_{pr} ) \over \sin(\pi \bar z_{pr} ) }
\right]\,,$$
when we neglect the singular parts at the coincident points.
As the result, the Schr\"odinger equation becomes
the free one, and   the transformed
wave-function (anyonic wavefunction) is multivalued
$$
\Phi^{({\rm anyon})} (1, \cdots, N) = \prod^N_{p>r}
\left(
{\sin(\pi z_{pr} )
\over \sin(\pi \bar z_{pr} ) }
\right)^{\nu \over 2}
\Phi (1, \cdots, N)\,,
$$
where $\nu = e^2 / (2\pi\hbar\mu c^2)$
and ${e \over \mu c} =\nu\phi_0$ with the unit flux
quantum $\phi_0 = hc/e$.

\section{Periodic property}

Until now we have considered  the theory on the fundamental
domain of the cylinder.
To find the boundary condition of the anyon equation,
we have to consider the theory on the covering
space, which consists of the repeated domains
of the fundamental one with the boundary
identified.
In this scheme,
the non-contractible loop on the cylinder
is identified with a line on the
fundamental domain from one edge to the other.
To describe the hamiltonian density
outside the fundamental domain,
let us denote $D_{m}$ for the domain with
$(x{'}^1 = x^1 + mL_1, x{'}^2 = x^2)$
where $x^i$ lies on the fundamental domain
$D_{0}$ and $m$ is an integer.
(In the following, we reserve the
unprimed coordinates for the ones in the fundamental domain
$D_{0}$ and primed for the domain $D_{m}$).

Obviously, the effective hamiltonian density
on $D_{m}$ can be written as
\begin{equation}
{\cal H}(x{'})
= {1 \over 2m} \left|
(i \hbar \partial_i - {e\over c} A_i^{[1]} (x{'})) \psi^{(2)} (x{'})
\right|^2\,,
\label{hamil1prime}
\end{equation}
where
$$A_i^{[1]} (x{'})
= - {e \over \mu c} \epsilon_{ij} \partial_{x^j}
\int \! d \vec y G_{\rm cyl}
(\vec x{'}, \vec y\,{'}) J_0(y{'})\,.$$
Noting the relation,
$G_{\rm cyl}(\vec x, \vec y) = G_{\rm cyl}(\vec x{'}, \vec y\,{'})$,
we have
\begin{equation}
A_i^{[1]} (x{'})
= - {e \over \mu c} \epsilon_{ij} \partial_{x^j}
\int \! d \vec y G_{\rm cyl}(\vec x, \vec y) J_0(y{'})\,.
\label{Ai1prime}
\end{equation}

This hamiltonian density on $D_{m}$ is canonically
related with the one on $D_{0}$
\begin{equation}
{\cal H}(x{'})
= T_{m} {\cal H}(x)  T^+_{m}\,,
\end{equation}
where
\begin{eqnarray}
J_0 (x{'}) &=& T_{m} J_0 (x) T^+_{m}\,,
\nonumber\\
b^{(2)}(x{'}) &=& T_{m} b^{(2)} (x) T^+_{m}\,,
\nonumber\\
\psi^{(2)}(x{'})
&=& T_{m} \psi^{(2)} (x) T^+_{m}\,.
\label{field1prime}
\end{eqnarray}
The explicit translation operator
along the non-contractible loop is given as
$T_{m} =  T_1^m$,
where
\begin{equation}
T_1 \equiv \exp -L_1 \int d\vec y \{\psi^{(2)+} (y)
\partial_{y^1} \psi^{(2)} (y)
+ i {\mu \over \hbar} a_1^{(2)} (y) \partial_{y^1} a_2^{(2)} (y) \}\,.
\end{equation}
We also note that the translation operator
connects the gauge field as
\begin{equation}
a_i^{(2)}(x{'}) = T_{m} a_i^{(2)} (x) T^+_{m}\,.
\label{aiprime}
\end{equation}

On the cylinder one can leave
the hamiltonian density $\cal H$
invariant under the translation along
the non-contractible loops as given in Eq. (\ref{field1prime})
if one defines the matter field as
$\psi^{(2)} (x{'}) = \psi^{(2)} (x) \exp iC_{m} $,
where $C_{m}$ is a constant on $D_{m}$.  We can
choose the constant $C_{m} = 0$ without losing any generality.
\begin{equation}
\psi^{(2)} (x{'}) = \psi^{(2)} (x) \,.
\label{psiperiod}
\end{equation}
This definition leads to the periodic condition
on $J_0$ and $b^{(2)}$ as
\begin{equation}
J_0 (x{'}) = J_0 (x) ; \quad
b^{(2)} (x{'}) = b^{(2)} (x)\,.
\label{J0period}
\end{equation}
This makes the effective gauge field
$A_i^{[1]} (x{'})$ manifestly
translation-invariant,
\begin{equation}
A_i^{[1]} (x{'}) = A_i^{[1]} (x)\,,
\label{Aperiod}
\end{equation}
and the translation-invariance of the hamiltonian density follows :
\begin{equation}
[T_m, {\cal H}(x)] = 0\,.
\label{THcomm}
\end{equation}

We note that the vacuum defined
on the fundamental domain remains the same on
the covering space due to Eq.(\ref{J0period}).
Therefore, the physical state is the same on all the
covering space and the Gauss constraint is identically satisfied.
The equal-time commutation relation
of $J_0(x)$ with $\psi^{(2)} (y{'})$ is
given as
\begin{equation}
[J_0(x), \psi^{(2)} (y') ] = [J_0(x), \psi^{(2)} (y) ]
=- \psi^{(2)} (y) \delta (\vec x - \vec y)\,.
\end{equation}
One can also define the periodic property
of the gauge field $a_i^{(2)}$ given in Eq.(\ref{aiprime}) up to a
gauge transformation,
which leaves the field strength $b^{(2)}$ invariant as
given in Eq.(\ref {J0period}),
\begin{equation}
a_i^{(2)} (x')
= a_i^{(2)} (x) + \partial_i \Omega (x)\,,  \label {aiperiod}
\end{equation}
where $\Omega(x)$ is an arbitrary periodic function.

The original fields
can be obtained using the canonical operator,
\begin{equation}
V_{2}^{(m)} (t)
= \exp i{e\over \hbar c} \int\! \int
d\vec x d \vec y J_0 (x{'} )
\left[ g_2(\vec x, \vec y)\theta_2(y')
+ G_p (\vec x, \vec y) \partial_k a_k^{(2)}(y{'} )
\right] \,,
\end{equation}
where the integration
$\int d\vec x$ denotes for
$\int_{0\le x^1<L_1, -\infty < x^2 < \infty } d \vec x$.
The operator is the same as in Eq. (\ref{V_2})
except that the field operators are replaced by the
ones on the domain  $D_{m}$.
The hamiltonian density becomes
\begin{equation}
{\cal H} (x{'}) = {1 \over 2m}
\left| (i \hbar \partial_i - {e\over c} a_i (x{'}))
\psi (x{'}) \right|^2\,, \label{hamilprime}
\end{equation}
where
\begin{eqnarray}
\psi(x{'})
&=& \exp -i {e \over \hbar c}
\int d \vec y
\left[ g_2(\vec x, \vec y) \theta_2 (y')
+ G_p (\vec x, \vec y)  \partial_{y^j} a_i^{(2)}(y{'})
\right] \psi^{(2)}(x{'})\,,
\nonumber\\
a_i(x{'})
&=& a_i^{(2)} (x{'})
- {e \over \mu c} \epsilon_{ij} \partial_{x^j}
\int d\vec y G_{\rm cyl}(\vec x, \vec y) J_0 (y)\,.
\label{psiprime}
\end{eqnarray}
The periodic property of the fields are easily deduced
using Eqs. (\ref {psiperiod}, \ref {aiperiod}),
\begin{eqnarray}
\psi (x{'}) &=& \psi(x) \exp - i {e \Omega (x) \over \hbar c}\,,
\nonumber\\
a_i (x') &=& a_i (x) + \partial_i \Omega (x)\,,  \label {fieldperiod}
\end{eqnarray}
which demonstrates that the periodic property
for $\psi(x)$ and $a_i(x)$ is
given as a local gauge transformation.
The commutation relations of the fields
are given as
$$ \{\psi(x{'}), \psi(y{''})\} = 0 $$
$$ \{\psi(x{'}), \psi^+(y{''})\} = \delta (\vec x - \vec y)$$
$$[a_1 (x{'}), a_2(y{''}) ]
= i {\hbar \over \mu}\delta (\vec x - \vec y)\,,$$
where $x'$ and $y''$ may lie in different domains.

Let us consider the many anyon wavefunctions
whose coordinates are located in different domains.
Similarly in Eq. (\ref {Nwave}),
we define the wavefunction as
\begin{equation}
\Phi (\{1\}, \cdots, \{N\})
\equiv < 0|\psi^{(2)} (\{1\}) \cdots \psi^{(2)} (\{N\}) |N > \,,
\label{Nwave1}
\end{equation}
where the coordinates
$\{i\} \equiv (x_1^{(i)} + m^{(i)} L_1, x_2^{(i)})$
with $m^{(i)}$ integer
may lie in any mixed domain.
The Schr\"odinger equation is the same
as the one given in Eq. (\ref {NSchp}) since
the hamiltonian density ${\cal H}(x{'})$
is translation-invariant :
\begin{equation}
i\hbar {\partial\Phi \over \partial t}
(\{1\}, \cdots, \{N\})
= {1 \over 2m} \sum^N_{p=1}
\{i\hbar \partial_i^{(p)} - {e  \over c}
\sum^N_{r=1(\ne p)}
{\cal A}_i (p,r)\}^2 \Phi(\{1\}, \cdots, \{N\}) \,.
\label{NSchp1}
\end{equation}
The hamiltonian has the periodic gauge-like potential
since it contains the
coordinate function of $x^{(i)} $ rather than $x^{(i)}{'} $.
One should note that the
periodic property of the matter field
in Eq. (\ref {psiperiod}) maintains the
trivially periodic wavefunction
\begin{equation}
\Phi (\{1\}, \cdots, \{N\}) =  \Phi (1, \cdots, N)\,,
\label{phiperiod}
\end{equation}
and the fermionic exchange property
\begin{equation}
\Phi (\cdots, \{i\}, \cdots, \{j\}, \cdots)
=  - \Phi (\cdots, \{j\}, \cdots, \{i\}, \cdots)\,.
\end{equation}

\section{Hard wall boundary condition}

The analysis for the torus and cylinder can be easily extended to
the box surrounded by the hard wall.
Let us consider the system in a box
($0 \le x^1 < L_1$
and $0 \le x^2 < L_2 $) with infinite potential barrier at the edge.
The gauge mode is expanded as
\begin{equation}
a_i(x) =
{2 \over \sqrt {L_1L_2 }}
\sum_{n_1, n_2 }
{1 \over { p^0}}
[ ( {p_i \over \mu} g(\vec p)
+ i{\epsilon_{ij} p_j \over 2  } h (\vec p))
\sin p_1x_1 \sin p_2x_2 e^{-ip^0 x^0}
+ c.c]_{p^0 = |\vec p|} \,,
\label{a1box}
\end{equation}
where $n_i = { L_i p_i \over \pi}$ is a positive  integer,
$g$ and $h$ satisfy
the discrete version of the commutation relation
given in Eq.~(\ref{ghcomm}).
$a_0(x)$ is expressed  as
\begin{equation}
a_0(x)
= \lambda  + {2 \over \sqrt {L_1L_2}}
\sum_{n_1,n_2 }
[ {g(\vec p)\over \mu}
\sin p_1x_1 \sin p_2x_2 e^{-ip^0 x^0}
+ c.c]_{p^0 = |\vec p|}\,.
\label{a0box}
\end{equation}

For the hamiltonian density $\cal H$ given
in Eq.~(\ref{hamil}),
we consider a canonical transformation operator
\begin{equation}
W(t)= \exp i
{e\over \hbar c}
\int\! \int d\vec x d \vec y
J_0^{(1)}(\vec x,t)
G_{\rm box}(\vec x, \vec y)
\partial_k a_k^{(1)}(\vec y, t) 
\,.
\label{W}
\end{equation}
$G_{\rm box}$ is the Green's function,
which vanishes at the edge of the box,
whose form is written as
\begin{equation}
G_{\rm box}(\vec x, \vec y)
= {1\over 4\pi} \ln
\left|
f(z,\bar z; \zeta, \bar\zeta )
\right|^2\,,
 \label{Greenbox}
\end{equation}
where
$z= (x^1+ i x^2 )/L_1$,
$\zeta =(y^1 + i y^2)/L_1$.
Explicitly,
\begin{equation}
f(z,\bar z; \zeta, \bar\zeta )
= {\theta_1 ({z-\zeta \over 2}|\tau)
\theta_1 ({z+\zeta \over 2} | \tau)
\over
\theta_1 ({z-\bar \zeta \over 2}|\tau)
\theta_1 ({z +\bar\zeta \over 2} | \tau)}
\end{equation}
where
$\theta_1$ is the odd Jacobi theta function
and $\tau = iL_2 /L_1$.
One may also express this Green's function in
terms of the Weierstrass' associated $\sigma$ function,
\begin{equation}
f(z,\bar z; \zeta, \bar\zeta )
= {\sigma ({z-\zeta })
\sigma ({z+\zeta })
\over
\sigma ({z-\bar \zeta })
\sigma ({z +\bar\zeta } )}\,.
\end{equation}

The original fields are given  in terms  of the new fields as
\begin{eqnarray}
\psi(x) & =& W(t) \psi^{(1)}(x) W^+(t)
\nonumber\\
& =& \psi^{(1)}(x)
\exp -i \{{e\over \hbar c}
\int \! d \vec y
G_{\rm box}(\vec x, \vec y)  \partial_{y^i} a_i^{(1)} (y) \}\,,
\nonumber\\
a_i(x) &=&
W(t) a_i^{(1)}(x) W^+(t)
\nonumber\\
&=& a_i^{(1)}(x) -
{e \over \mu c}
\epsilon_{ij} \partial_{x^j}
\int \! d\vec y {G}_{\rm box}(\vec x, \vec y) J_0^{(1)} (y)\,,
\nonumber\\
\label{newfield}
\end{eqnarray}
where
$J_0(x)= J_0^{(1)}(x)$.

The hamiltonian density ${\cal H}$
becomes of the form in Eq.~(\ref{hamil1})
and the effective gauge field is given as
\begin{equation}
a_i^{({\rm eff})}(x)
= \epsilon_{ij} \partial_{x^j}
\int \! d \vec y {G}_{\rm box}(\vec x, \vec y) (b^{(1)}(y)
-{e \over \mu c} J_0(y))
\,. \label{aeff1}
\end{equation}
As done in section 2, we may drop  $b^{(1)}(x)$ completely
since it generates the gauge transformation and commutes with
other modes.
Therefore, the hamiltonian density becomes,
\begin{equation}
{\cal H}(x)
=  {1 \over 2m} \left| (i \hbar \partial_i -
{e\over c} A_i^{[1]}(x) )\psi^{(1)}(x) \right|^2\,,
\label{hamilbox}
\end{equation}
where
\begin{equation}
A_i^{[1]}(x)= - {e \over \mu c}
\epsilon_{ij} \partial_{x^j}
\int \! d \vec y {G}_{\rm box}(\vec x, \vec y) J_0(y)\,.
\label{Aib}
\end{equation}

Now the many-anyon equation can be obtained easily.
We have only to  replace the Green's function in Eq.~(\ref{Kgreencyl})
by the one given in Eq.~(\ref{Greenbox}).
The Aharonov-Bohm potential appears adapted to the boundary condition,
which can be transformed away
leaving the multi-valued wavefunction as
$$
\Phi^{({\rm anyon})} (1, \cdots, N) = \prod^N_{p>r}
\left({f(z_p, \bar z_p; z_r, \bar z_r)
\over f^*(z_p, \bar z_p; z_r, \bar z_r)
} \right)^{\nu \over 2}
\Phi (1, \cdots, N)\,.$$

Finally, one may also apply the theory inside strip
with the  hard-wall at edge. In this case the Green's
function becomes of the form
\begin{equation}
G_{\rm strip}(\vec x, \vec y) =
{1 \over 4\pi} \ln
|F(z, \zeta, \bar \zeta)|^2\,,
\nonumber
\end{equation}
where
$$
F(z, \zeta, \bar \zeta) = \prod_{n= - \infty}^\infty
\left(
{z-\zeta -2n \over z+\bar \zeta -2n}
\right)\,.
$$
 From this one can obtain easily the anyon equation
following the above analysis.

\section{Summary and Discussion}

We have analyzed the Chern-Simons theory
coupled to non-relativistic matter
field on a cylinder in analogy with the case
for the infinite plane and on the torus.
Quantizing the field first and performing canonical
transformation we obtain
an effective theory in terms of matter field
such that it manifests the gauge invariance and the
translation invariance along the non-contractible loop.
The periodic property of the original field along the
non-contractible loop is determined up to the
gauge transformation, which is demonstrated
from the simple periodic property of the
new fields
through the canonical transformation.
The canonical transformation operator
can be constructed
either like in the infinite plane case
$V_1(t)$ in
Eq. (\ref{V_1})
or in the torus case
$V_2(t)$ in Eq. (\ref{V_2}),
which result in the same effective hamiltonian
and anyon equation.

Unlike on a torus
the gauge field on the cylinder in Eq.~(\ref{aicyl})
does not need the extra mean magnetic field contribution
since both the mean field contribution and
the analogue of the zero-mode
$\theta_i(x)$  as
given in Eq.~(\ref{thetamode})
are contained as
necessary ingredients to satisfy the commutation relation
given in Eq. (\ref{commfield}).
This mean field contribution has the important
role to decouple the gauge field
in the canonical transformation.
The typical physical state constructed as
$$ |N > \sim \int d \vec x^{(1)}
\cdots  d \vec x^{(N)} \psi^{(2)+} (1) \cdots
\psi^{(2)+} (N) | 0 >\,, $$
($\psi^{(2)+}$ is the gauge invariant field defined
in Eq. (\ref{field2def}))
satisfies the Gauss constraint automatically
and flux quantization is not needed
since
 $\int d\vec y b(y) |N > =  - {e \over \mu c} N|N > $
 due to the mean field contribution in the gauge field and
$Q|N > = N|N > $ where $Q = \int d \vec x J_0(x)$.

One may take the Wilson loop operator into consideration,
which is defined as
$W(t)= \exp i\int_{-\infty}^{\infty} dy
\theta_1^{(2)}(x,y,t) \kappa(y)$,
where $\kappa(y)$ is an arbitrary distribution.
One can easily check that $W$ is gauge invariant
and is a  constant of motion,
$$
[W, \Gamma ] = 0; \quad
[W, H]=0.
$$
Therefore, one may define the wavefunction as
$$
\Phi^{(w)} (1, \cdots, N)
\equiv < 0|W \psi^{(2)} (x^{(1)}) \cdots \psi^{(2)} (x^{(N)}) |N > \,.
$$
instead of the definition given in Eq. (\ref{Nwave}).
This newly-defined  wavefunction is also gauge invariant
and satisfies the same Schr\"odinger equation in
Eq. (\ref{NSchp}).
It is obvious that this wavefunction has the
simple periodic condition as given in
Eq. (\ref{phiperiod}).

We also considered the system inside a box
with the hard wall boundary condition, where
the zero-modes do not appear.
There are mean magnetic field
modes in the gauge field, which
play the similar role of $\theta_i$ on the cylinder
and decouple the gauge fields
from the physical states
after the canonical transformation.
The canonical transformation operator is similar to the
one given in the  plane case.
The resulting many-body Schr\"odinger
equation contains the
Aharonov-Bohm potential adapted to
the hard-wall boundary condition.

Finally, we note that
even though the many-body equation looks very simple
(it becomes free equation when the potential is
singular gauged away), its solution is not
easy to get at.
In this situation,
to understand the clear connection
between the degenerate ground
states of the anyons on the infinite plane with
magnetic field and
the eigenstates of the Calogero model
\cite{vasil}
will be helpful
for obtaining the idea about
the spectrum and wavefunctions
of the many-body states
in general.

\bigskip
\centerline {\bf Acknowledgments}

This work is supported in part by KOSEF No. 931-0200-030-2
and by SRC program through Seoul National University.


\bigskip
\begin {thebibliography}{99}

\bibitem {deser}  S. Deser, R. Jackiw and S. Templeton,
Ann. Phys. (N. Y.)  {\bf 140} (1982) 372.
\bibitem {hagen}  C. Hagen, Ann. Phys. (N. Y.)
{\bf 157} (1984) 342;
Phys. Rev. D {\bf 31} (1985) 2135.
\bibitem {goldhaber}  A. S. Goldhaber, R. Mackenzie and F. Wilczek,
Mod. Phys. Lett. A {\bf 4} (1989) 21;
T. H. Hansson, M. Ro\u{c}ek, I. Zahed and S. C. Zhang,
Phys. Lett. {\bf B214} (1988) 475.
\bibitem {semenoff}  G. W. Semenoff,
Phys. Rev. Lett. {\bf 61} (1988) 517;
G. W. Semenoff, P. Sodano and Y. S. Wu,
{\it ibid}. {\bf 62} (1989) 715;
G. W. Semenoff and P. Sodano,
Nucl. Phys. B {\bf 328} (1989) 753.
\bibitem {matsuyama}  T. Matsuyama,
Phys. Lett. {\bf B228} (1989) 99;
Phys. Rev. D {\bf 15} (1990) 3469;
D. Boyanovsky, E. Newman and C. Rovelli, Phys.
Rev. D {\bf 45} (1992) 1210;
R. Banerjee, Phys. Rev. Lett. {\bf 69} (1992) 17;
Nucl. Phys. B {\bf 390} (1993) 681.
\bibitem {swanson-cho}   M. S. Swanson,
Phys. Rev. D {\bf 42} (1990) 552;
K. H. Cho and C. Rim,
Int. J. Mod. Phys. A {\bf 7} (1992) 381.
\bibitem {leinaas}  J. Leinaas and J. Myrheim,
Nuovo Cim. B {\bf 37} (1977) 1;
G. Goldin, R. Menikoff and D. H. Sharp,
J. Math. Phys. {\bf 22} (1981) 1664;
F. Wilczeck, Phys. Rev. Lett. {\bf 49} (1982) 957;
Y. S. Wu, Phys. Rev. Lett. {\bf 52} (1984) 2103.
\bibitem {jackiw}  R. Jackiw and S. -Y. Pi,
Phys. Rev. D {\bf 42} (1990) 3500.
\bibitem {kimlee}  C. Kim, C. Lee, P. Ko, B. -H. Lee
and H. Min, Phys. Rev. D
{\bf 48} (1993) 1821.
\bibitem {randjbar}  S. Randjbar-Daemi, A. Salam and J. Strathdee,
Phys. Lett. {\bf B240} (1990) 121;
R. Iengo and K. Lechner,
Nucl. Phys. B {\bf 346} (1990) 551,
{\it ibid}. {\bf 364}, 551 (1991);
Y. Hosotani, Phys. Rev. Lett.
{\bf 62} (1989) 2785;
C. -L. Ho and Y. Hosotani,
Int. J. Mod. Phys. A {\bf 7} (1992) 5797;
Phys. Rev. Lett. 70 (1993) 1360;
A. Fayyazuddin,
Nucl. Phys. B {\bf 401} (1993) 644.
\bibitem{hoso} S. Chakravarty and Y. Hosotani,
Phys. Rev. D {\bf 44} (1991) 441.
\bibitem{rim-torus} K.-H. Cho and C. Rim,
preprint SNUTP-93-96/hepth-9312204.
\bibitem {polychro}  A. P. Polychronakos,
Ann. Phys. {\bf 203} (1990) 231.
\bibitem{vasil} L. Brink, T. H. Hansson, S. Konstein and M. A. Vasiliev,
Nucl. Phys. B {\bf 401} (1993) 599.

\end {thebibliography}
\end{document}